\documentclass[12pt]{article}
\usepackage[table]{xcolor}
\usepackage{colortbl}
\definecolor{lightgray}{gray}{0.9}
\usepackage{graphicx}
\usepackage{lscape}
\usepackage{citesort}
\usepackage{amssymb}
\usepackage{appendix}
\usepackage{multirow}

\usepackage{graphicx}
\usepackage{lscape}
\usepackage{citesort}
\usepackage{amssymb}
\usepackage{appendix}
\usepackage{multirow}

\begin{document}
\def\qq{\langle \bar q q \rangle}
\def\uu{\langle \bar u u \rangle}
\def\dd{\langle \bar d d \rangle}
\def\sp{\langle \bar s s \rangle}
\def\GG{\langle g_s^2 G^2 \rangle}
\def\Tr{\mbox{Tr}}
\def\figt#1#2#3{
        \begin{figure}
        $\left. \right.$
        \vspace{-2cm}
        \begin{center}
        \includegraphics[width=10cm]{#1}
        \end{center}
        \vspace{-0.2cm}
        \caption{#3}
        \label{#2}
        \end{figure}
    }

\def\figb#1#2#3{
        \begin{figure}
        $\left. \right.$
        \vspace{-1cm}
        \begin{center}
        \includegraphics[width=10cm]{#1}
        \end{center}
        \vspace{-0.2cm}
        \caption{#3}
        \label{#2}
        \end{figure}
                }

\def\ds{\displaystyle}
\def\beq{\begin{equation}}
\def\eeq{\end{equation}}
\def\bea{\begin{eqnarray}}
\def\eea{\end{eqnarray}}
\def\beeq{\begin{eqnarray}}
\def\eeeq{\end{eqnarray}}
\def\ve{\vert}
\def\vel{\left|}
\def\ver{\right|}
\def\nnb{\nonumber}
\def\ga{\left(}
\def\dr{\right)}
\def\aga{\left\{}
\def\adr{\right\}}
\def\lla{\left<}
\def\rra{\right>}
\def\rar{\rightarrow}
\def\lrar{\leftrightarrow}
\def\nnb{\nonumber}
\def\la{\langle}
\def\ra{\rangle}
\def\ba{\begin{array}}
\def\ea{\end{array}}
\def\tr{\mbox{Tr}}
\def\ssp{{\Sigma^{+}}}
\def\sso{{\Sigma^{0}}}
\def\ssm{{\Sigma^{-}}}
\def\xis0{{\Xi^{0}}}
\def\xism{{\Xi^{-}}}
\def\qs{\la \bar s s \ra}
\def\qu{\la \bar u u \ra}
\def\qd{\la \bar d d \ra}
\def\qq{\la \bar q q \ra}
\def\gGgG{\la g^2 G^2 \ra}
\def\q{\gamma_5 \not\!q}
\def\x{\gamma_5 \not\!x}
\def\g5{\gamma_5}
\def\sb{S_Q^{cf}}
\def\sd{S_d^{be}}
\def\su{S_u^{ad}}
\def\sbp{{S}_Q^{'cf}}
\def\sdp{{S}_d^{'be}}
\def\sup{{S}_u^{'ad}}
\def\ssp{{S}_s^{'??}}

\def\sig{\sigma_{\mu \nu} \gamma_5 p^\mu q^\nu}
\def\fo{f_0(\frac{s_0}{M^2})}
\def\ffi{f_1(\frac{s_0}{M^2})}
\def\fii{f_2(\frac{s_0}{M^2})}
\def\O{{\cal O}}
\def\sl{{\Sigma^0 \Lambda}}
\def\es{\!\!\! &=& \!\!\!}
\def\ap{\!\!\! &\approx& \!\!\!}
\def\md{\!\!\!\! &\mid& \!\!\!\!}
\def\ar{&+& \!\!\!}
\def\ek{&-& \!\!\!}
\def\kek{\!\!\!&-& \!\!\!}
\def\cp{&\times& \!\!\!}
\def\se{\!\!\! &\simeq& \!\!\!}
\def\eqv{&\equiv& \!\!\!}
\def\kpm{&\pm& \!\!\!}
\def\kmp{&\mp& \!\!\!}
\def\mcdot{\!\cdot\!}
\def\erar{&\rightarrow&}
\def\olra{\stackrel{\leftrightarrow}}
\def\ola{\stackrel{\leftarrow}}
\def\ora{\stackrel{\rightarrow}}

\def\simlt{\stackrel{<}{{}_\sim}}
\def\simgt{\stackrel{>}{{}_\sim}}


\title{
         {\Large {\bf Transition Form Factors of  $\chi_{b2}(1P)\rightarrow B_{c}\bar{l}\nu$ in QCD}}}
\author {\small K. Azizi$^{1\, \dag}$,  H. Sundu$^{2\, \ddag}$,  J.
Y. S\"{u}ng\"{u}$^{2\, \ast}$, N. Yinelek$^{2\,\ast\ast}$ \\
\small $^1$Department of Physics, Dogus University, Acibadem-Kadikoy, 34722 Istanbul, Turkey\\
$^2$\small Physics Department, Kocaeli University, 41380 Izmit, Turkey\\
$^\dag$\small e-mail:kazizi@dogus.edu.tr \\
$^\ddag$\small e-mail:hayriye.sundu@kocaeli.edu.tr \\
$^\ast$\small e-mail:jyilmazkaya@kocaeli.edu.tr \\
$^{\ast\ast}$\small e-mail:neseyinelek@gmail.com}
\date{}
\begin{titlepage}
\maketitle
\thispagestyle{empty}
\begin{abstract}
The form factors of the semileptonic $\chi_{b2}(1P)\rightarrow
B_{c}\bar{l}\nu$ decay are calculated  using the QCD sum rule
approach. The results obtained are then used to estimate the
decay widths of this transition in all lepton channels. The orders
of decay rates indicate that this transition is accessible at LHC
for all lepton channels.
\end{abstract}
~~~~~~~~PACS number(s):11.55.Hx, 14.40.Pq, 13.20.He
\end{titlepage}

\section{Introduction}

Quarkonia are flavorless bound states composed of combinations
of quarks and their antiquarks. Since discovery of $J/\psi$ meson
in 1974, many new quarkonia states have been
detected. The quarkonia systems consist of charmonium and
bottomonium. Due to the large mass there are no toponium bound
states and no light quark-antiquark states because of the mixture
of the light quarks in experiments. The large mass difference
between charm and bottom quarks prevents them from mixing.
The heavy quark bound states may provide  key tools for understanding  the
interactions between quarks, new hadronic production mechanisms
and transitions, the magnitude of the CKM matrix elements and also
analyzing the results of heavy-ion experiments.

The  $J/\Psi$ suppression in ultra relativistic heavy-ion
collisions was first suggested as a signal of the formation of a quark-gluon plasma (QGP).
However, most recently, attentions have shifted to the bottomonium states
due to that they are more massive than charmonium states. The
bottom quarks and anti-quarks are relatively rare within the
plasma, so the probability for reproduction of the bottomonium
states through recombination is much smaller than for charm
quarks. Consequently, it is expected that the bottomonium system
to be a cleaner probe of the QGP than the charmonium system. Hence
investigations on the properties of bottomonium systems can help
us get useful information not only about the nature of the
$b\bar{b}$ systems, but also on the existence of QGP.

 Since the discovery of quarkonium states, QCD
sum rule technique as one of the most
powerful nonperturbative tools to hadron physics \cite{Shifman1,Shifman2} has played an important role in understanding
the quarkonia spectrum.  In order to find missing states one should
know their physical properties to develop a successful search
strategy. Clearly significant progress in understanding of
quarkonium production cannot be reached without detailed
measurements of the cross sections and fractions of the quarkonia.
Thereby, form factors and decay widths of quarkonia become
significant for completing quarkonia spectrum. Completing the
bottomonium spectrum is a crucial validation of theoretical
calculations and a test of our understanding of bottomonium states
in the context of the quark model. Bottomonium states are
considered as great laboratories to search for the properties od QCD at low energies. 

In this connection, we investigate the decay properties of tensor
$\chi_{b2}(1P)$  meson as one of the important members of the bottomonia to the heavy $B_{c}$ meson in the present work.
The $B_{c}$ meson with $J^{P} = 0^{-}$ is the only meson consists
of two heavy quarks with different flavors. Yet, other possible
$B_{c}$ states (the scalar, vector, axial-vector and tensor) have
not been observed, however many new $B_{c}$ species are expected
to be produced at the Large Hadron Collider (LHC) in the near
future.

Taking into account the two-gluon condensate corrections, the
transition form factors of the semileptonic
$\chi_{b2}(1P)\rightarrow B_{c}\bar{l}\nu$ decay channel are
calculated within the three-point QCD sum rule.
We use the values of transition form factors to estimate the
decay rate of the transition under consideration at all lepton
channels. The interpolating current of $\chi_{b2}(1P)$ with
quantum numbers $I^{G}(J^{PC}) = 0^{+}(2^{++})$ contains
derivatives with respect to space-time. So, we start our
calculations in the coordinate space then we apply the Fourier
transformation to go to the momentum space. To suppress the
contributions of the higher states and continuum, we apply a
double Borel transformation.

The article is organized as follows. We derive the QCD sum rules
for the transition form factors in Section II. Last section is
devoted to the numerical analysis of the obtained sum rules, estimation of the decay rates at all lepton channels and  concluding remarks.
\section{QCD Sum Rules for $\chi_{b2}\rightarrow B_{c}\bar{l}\nu$ Transition Form Factors}
The semileptonic $\chi_{b2}\rightarrow B_{c}\bar{l}\nu$ decay is
based on $b\rightarrow c\bar{l}\nu$ transition at quark level whose
effective Hamiltonian can be written as
\begin{equation}\label{Hamiltonian}
{\cal H}_{eff}(b \rightarrow c\bar l \nu_{l}) = \frac{G_F}{\sqrt2}
V_{cb} ~\overline{c} \gamma_\mu (1-\gamma_5) b l \gamma^{\mu}
(1-\gamma_5){\overline{\nu}},
\end{equation}
where $G_F$ is the Fermi weak coupling constant and $V_{cb}$ is
element of the Cabibbo-Kobayashi-Maskawa (CKM) mixing matrix.
After sandwiching the effective Hamiltonian between the initial
and final states, the amplitude of this transition is obtained in
terms of transition matrix elements. These matrix elements will be
parameterized in terms of transition form factors later.

In order to start our calculations, we consider the three-point
correlation function
\begin{eqnarray}\label{correl.func.}
\Pi_{\mu\alpha\beta}(p,p^{\prime},q)=i^2\int d^{4}x e^{-ip\cdot
x}\int d^{4}y e^{ip'\cdot y}\langle0|{\cal
T}|{j_{B_c}(y)j_\mu^{tr,V-A}(0)j^{\dag\chi_{b2}}_{\alpha\beta}(x)}|0\rangle~,
\end{eqnarray}
where ~${\cal T}$ is the time-ordering operator and $j_{\mu}^{tr,
V-A}(0)=\bar c(0) \gamma_\mu(1-\gamma_5)
 b(0)$ is the transition current. To proceed we also
need the interpolating currents of the initial and final mesons in
terms of the quark fields, which are given as
\begin{equation}\label{tensorcurrent}
j _{\alpha\beta}^{\chi_{b2}}(x)=\frac{i}{2}\left[\bar b(x)
\gamma_{\alpha} \olra{\cal D}_{\beta}(x) b(x)+\bar b(x)
\gamma_{\beta}  \olra{\cal D}_{\alpha}(x) b(x)\right]
\end{equation}
and
\begin{equation}\label{tensorcurrent}
j_{B_{c}}(y)=\bar b(y)\gamma_5c(y),
\end{equation}
where the covariant derivative $\olra{\cal D}_{\beta}(x)$ denotes
the four-derivative with respect to $x$ acting on two sides
simultaneousely and it is defined as
\begin{equation}\label{derivative}
\olra{\cal D}_{\beta}(x)=\frac{1}{2}\left[\ora{\cal D}_{\beta}(x)-
\ola{\cal D}_{\beta}(x)\right],
\end{equation}
with
\begin{eqnarray}\label{derivative2}
\overrightarrow{{\cal
D}}_{\beta}(x)=\overrightarrow{\partial}_{\beta}(x)-i
\frac{g}{2}\lambda^aA^a_\beta(x)
\end{eqnarray}
and
\begin{eqnarray}\label{derivative3}
\overleftarrow{{\cal
D}}_{\beta}(x)=\overleftarrow{\partial}_{\beta}(x)+
i\frac{g}{2}\lambda^aA^a_\beta(x).
\end{eqnarray}
Here,  $\lambda^a$ are the Gell-Mann matrices and $A^a_\beta(x)$
denote the external  gluon fields.

According to the method used, the correlation function in
Eq.(\ref{correl.func.}) is calculated in two different
ways. In physical or phenomenological side we obtain
it in terms of hadronic parameters such as masses and decay
constants. In QCD or theoretical side we evaluate it in terms of
QCD degrees of freedom like quark masses as well as quark and
gluon condensates via operator product expansion (OPE). The QCD
sum rules for form factors are obtained by equating the above
representations to each other. After applying a double Borel
transformation, the  contributions of the higher states and continuum
are suppressed.

The hadronic side of the correlation function is obtained by
inserting complete sets of intermediate states into
Eq.(\ref{correl.func.}). After performing the four-integrals over
$x$ and $y$, we get

\begin{eqnarray} \label{phys1} \Pi_{\mu\alpha\beta}^{PHYS}(p,p',q)&=&\frac{\langle 0 \mid
 j_{B_{c}}(0)\mid
B_{c}(p') \rangle \langle B_{c}(p') \mid
j_\mu^{tr,V-A}\mid\chi_{b2}(p,\varepsilon) \rangle \langle
\chi_{b2}(p,\varepsilon)\mid
j^{\dag\chi_{b2}}_{\alpha\beta}(0)\mid
0\rangle}{(p'^2-m_{B_{c}}^2)(p^2-m_{\chi_{b2}}^2)}\nonumber
\\
&+&\cdots~,
\end{eqnarray}
where $\cdots$ denotes the contribution of the higher states and
continuum. To go further, we need to know the following matrix
elements:
\begin{equation}\label{matrixel1a} \langle B_{c}(p') \mid  j_{\mu}^{tr,V} \mid \chi_{b2}(p, \varepsilon) \rangle= h(q^2)\epsilon_{\mu\nu\theta\eta}
\epsilon^{\nu\lambda}P_{\lambda}P^{\theta}q^{\eta}~,
\end{equation}
\begin{equation}\label{matrixel1b} \langle B_{c}(p') \mid  j_{\mu}^{tr,A} \mid \chi_{b2}(p,\varepsilon) \rangle= -i \Big\{K(q^2)\epsilon_{\mu\nu}P^{\nu} +
\epsilon_{\theta\eta}P^{\theta}P^{\eta}[P_{\mu}b_{+}(q^2)+q_{\mu}b_{-}(q^2)]\Big\}~,
\end{equation}
\begin{equation}\label{kayb2current}
\langle \chi_{b2}(p,\varepsilon) \mid
j^{\dag\chi_{b2}}_{\alpha\beta}\mid
0\rangle=f_{\chi_{b2}}m_{\chi_{b2}}^3\varepsilon_{\alpha\beta}^*
\end{equation}
and
\begin{equation}\label{eq:Bccurrent}
\langle 0\mid j_{B_c}\mid B_{c}(p') \rangle=i \frac{f_{B_{c}}
m_{B_{c}}^2}{m_c+m_b},
\end{equation}
where $h(q^2)$, $K(q^2)$, $b_{+}(q^2)$ and $b_{-}(q^2)$ are
transition form factors, $\epsilon_{\alpha\beta}$ is the
polarization tensor associated with the $\chi_{b2}$ tensor meson;
$f_{\chi_{b2}}$ and $f_{B_{c}}$ are leptonic decay constants of
$\chi_{b2}$ and $B_{c}$ mesons, respectively,
$P_\mu=(p+p')_{\mu}$ and $q_{\mu}=(p-p')_{\mu}$.

Putting all matrix elements given in Eqs.~(\ref{matrixel1a}),
(\ref{matrixel1b}), (\ref{kayb2current}) and (\ref{eq:Bccurrent})
into Eq.(\ref{phys1}), the final representation of the correlation
function on physical side is obtained as
\begin{eqnarray}\label{phen2}
\Pi_{\mu\alpha\beta}^{PHYS}(p,p',q)&=&
\frac{f_{\chi_{b2}}f_{B_{c}} m_{\chi_{b2}} m_{B_{c}}^2}
{8(m_b+m_c)(p^2-m_{\chi_{b2}}^2)(p'^2-m_{B_{c}}^2)}\Big\{\Delta
K(q^2)q_{\alpha}g_{\beta\mu}\nonumber \\
&-&\frac{2}{3}\Big[\Delta'b_{-}(q^2)+\Delta K(q^2)\Big]q_{\mu}g_{\alpha\beta} \nonumber \\
 &-& \frac{2}{3}\Big[\Delta'b_{+}(q^2)+K(q^2)(\Delta+4m_{\chi_{b2}}^2)\Big ]P_{\mu}g_{\alpha\beta} \nonumber \\
 &-& i(\Delta-4m_{\chi_{b2}}^2) h(q^2)\varepsilon_{\lambda\eta\beta\mu}P_{\lambda}P_{\eta}q_{\alpha}+\mbox{other structures}\Big \}+...,
\end{eqnarray}
where
\begin{eqnarray}\label{delta1}
\Delta &=&m_{B_{c}}^2+3m_{\chi_{b2}}^2-q^2
\end{eqnarray}
and
\begin{eqnarray}\label{delta2}
\Delta'&=&m_{{B_c}}^4-2m_{{B_c}}^2(m_{\chi_{b2}}^2+q^2)+(m_{\chi_{b2}}^2-q^2)^2.
\end{eqnarray}
Note that we represented only the structures which we will use to find the
corresponding form factors. Meanwhile, we used the following
summation over the polarization tensors to obtain
Eq.~(\ref{phen2})
\begin{equation}\label{polarizationt1}
\sum_{\lambda}\varepsilon_{\mu\nu}^{\lambda}\varepsilon_{\alpha\beta}^{*\lambda}=\frac{1}{2}\eta_{\mu\alpha}\eta_{\nu\beta}+
\frac{1}{2}\eta_{\mu\beta}\eta_{\nu\alpha}
-\frac{1}{3}\eta_{\mu\nu}\eta_{\alpha\beta},
\end{equation}
where
\begin{equation}\label{polarizationt2}
\eta_{\mu\nu}=-g_{\mu\nu}+\frac{p_\mu p_\nu}{m_{\chi_{b2}^2}}.
\end{equation}
The next step is to calculate the QCD side of the correlation
function in deep Euclidean region, where $p^2\rightarrow-\infty$
and $p'{^2}\rightarrow-\infty$ via OPE. Placing the explicit
expressions of the interpolating currents into the correlation
function and contracting out all quark pairs via Wick's theorem,
we obtain
\begin{eqnarray}\label{correl.func.2}
\Pi^{QCD}_{\mu\alpha\beta}(p,p',q)&=&\frac{-i^3}{2}\int d^{4}x\int
d^{4}y e^{-ip\cdot x}e^{ip'\cdot y}
\nonumber \\
&\times& \Bigg\{Tr\left[S_b^{ia}(x-y)\gamma_5
S_c^{aj}(y)\gamma_\mu(1-\gamma_5)\olra{\cal
D}_{\beta}(x)S_b^{ji}(-x)\gamma_\alpha\right]+\left[\beta\leftrightarrow\alpha\right]\Bigg\},\nonumber\\
\end{eqnarray}
where $S$ is the heavy quark propagator and it is given by
\cite{Reinders85}
\begin{eqnarray}
S_{Q}^{ai}(x)&=&\frac{i}{(2\pi)^4}\int d^4k e^{-ik \cdot (x)}
\left\{ \frac{\delta_{ai}}{\!\not\!{k}-m_Q}
-\frac{g_sG^{\psi\varphi}_{ai}}{4}\frac{\sigma_{\psi\varphi}(\!\not\!{k}+m_Q)+
(\!\not\!{k}+m_Q)\sigma_{\psi\varphi}}{(k^2-m_Q^2)^2}\right.\nonumber \\
&+& \left.\frac{\pi^2}{3} \langle \frac{\alpha_sGG}{\pi}\rangle
\delta_{ai}m_Q \frac{k^2+m_Q\!\not\!{k}}{(k^2-m_Q^2)^4}
+\cdots\right\},
\end{eqnarray}
where $Q=b$ or $c$ quark. Replacing the explicit expression of the
heavy quark propagators in Eq.~(\ref{correl.func.2}) and applying
integrals over $x$ and $y$, we find the QCD side as
\begin{eqnarray}\label{eq:QCDside}
\Pi^{QCD}_{\mu\alpha\beta}(p,p',q)
&=&\Big(\Pi^{pert}_1(q^2)+\Pi^{nonpert}_1(q^2)\Big)q_{\alpha}g_{\beta\mu}+
\Big(\Pi^{pert}_2(q^2)+\Pi^{nonpert}_2(q^2)\Big)q_{\mu}g_{\beta\alpha}\nonumber \\
&+&
\Big(\Pi^{pert}_3(q^2)+\Pi^{nonpert}_3(q^2)\Big)P_{\mu}g_{\beta\alpha}+
\Big(\Pi^{pert}_4(q^2)+\Pi^{nonpert}_4(q^2)\Big)\varepsilon_{\lambda\nu\beta\mu}P_{\lambda}
P_{\alpha}q_{\nu}
\nonumber \\
&+&\mbox{other\, structures}.
\end{eqnarray}
Here $\Pi^{pert}_i(q^2)$ with $i=1,2,3,4$ are the perturbative
parts which are expressed in terms of  double dispersion integrals
as
\begin{eqnarray} \label{QCDside1}
\Pi^{pert}_i(q^2)=\int^{}_{}ds\int^{}_{}ds'
\frac{\rho_i(s,s',q^2)}{(s-p^2)(s'-p'^2)}+\mbox{subtracted terms},
\end{eqnarray}
where the spectral densities are defined as
$\rho_i(s,s',q^2)=\frac{1}{\pi}Im[\Pi^{pert}_i]$. The spectral
densities corresponding to  four different structures shown in
Eq.~(\ref{eq:QCDside}) are obtained as
\begin{eqnarray}\label{rho}
\rho_1(s,s',q^2)&=& \int_{0}^{1}dx
\int_{0}^{1-x}dy~~\frac{3\Big[m_c(-3+4x+2y)+m_b(-5+8x+4y)\Big]}{16\pi^2},
\nonumber \\
\rho_2(s,s',q^2)&=& \int_{0}^{1}dx \int_{0}^{1-x}dy~~\frac{
3\Big[m_c(3-4x-2y)+m_b(-1+4x+2y)\Big]}{8\pi^2},
\nonumber \\
\rho_3(s,s',q^2)&=& -\int_{0}^{1}dx
\int_{0}^{1-x}dy~~\frac{3\Big[m_c(1-2y)+m_b(1+2y)\Big]}{8\pi^2}
\end{eqnarray}
and
\begin{eqnarray}\label{rho2}
\rho_4(s,s',q^2)=0.
\end{eqnarray}
The function $\Pi_i^{nonpert}(q^2)$ in nonperturbative parts are
calculated from a similiar manner and by considering the two-gluon condensates contributes. Having calculated both the physical and
OPE sides of the correlation function, now, we match them to find
QCD sum rules for form factors.
In the Borel scheme we get
\begin{eqnarray}\label{K}
K(q^2)&=&\frac{8(m_b+m_c)}{f_{\chi_{b2}}f_{B_{c}} m_{B_{c}}^2
m_{\chi_{b2}}{\Delta}}e^{{m_{\chi_{b2}}^2}/{M^2}}e^{{m_{B_{c}}^2}/{M'^{2}}}
\nonumber \\
&\times&\Bigg\{\int^{s_0}_{4m_b^2}ds
\int^{s_{0}^{'}}_{(m_b+m_c)^2}ds'~~
e^{\frac{-s}{M^2}}e^{\frac{-s'}{M'^2}}\rho_1(s,s',q^2)\Theta[L(s,s',q^2)]
+\widehat{{\cal B}}_{M^2}\widehat{{\cal
B}}_{M'^2}\Pi_1^{nonpert}(q^2)\Bigg\},
\nonumber \\
b_{-}(q^2)&=&-\frac{12(m_b+m_c)}{f_{\chi_{b2}}
f_{B_{c}}m_{B_{c}}^2
m_{\chi_{b2}}{\Delta'}}e^{{m_{\chi_{b2}}^2}/{M^2}}
e^{{m_{B_{c}}^2}/{M'^2}}
\nonumber \\
&\times&\Bigg\{\int^{s_0}_{4m_b^2}ds
\int^{s_{0}^{'}}_{(m_b+m_c)^2}ds'
e^{\frac{-s}{M^2}}e^{\frac{-s'}{M'^2}}\rho_2(s,s',q^2)\Theta[L(s,s',q^2)]+\widehat{{\cal
B}}_{M^2}\widehat{{\cal B}}_{M'^2}\Pi_2^{nonpert}(q^2)
\nonumber\\
&-&\frac{\Delta}{\Delta'}K(q^2)\Bigg\},
\nonumber \\
b_{+}(q^2)&=&-\frac{12(m_b+m_c)}{f_{\chi_{b2}}f_{B_{c}}m_{B_{c}}^2m_{\chi_{b2}}{\Delta'}}e^{{m_{\chi_{b2}}^2}/{M^2}}e^{{m_{B_{c}}^2}/{M'^2}}
\nonumber \\
&\times&\Bigg\{\int^{s_0}_{4m_b^2}ds
\int^{s_{0}^{'}}_{(m_b+m_c)^2}ds'
e^{\frac{-s}{M^2}}e^{\frac{-s'}{M'^2}}\rho_3(s,s',q^2)\Theta[L(s,s',q^2)]+\widehat{{\cal
B}}_{M^2}\widehat{{\cal B}}_{M'^2}\Pi_3^{nonpert}(q^2)
\nonumber \\
&-&\frac{\Delta+4m_{\chi_{b2}}^2}{\Delta'}K(q^2)\Bigg\}
\nonumber \\
\end{eqnarray}
and
\begin{eqnarray}\label{h}
h(q^2)&=&-i\frac{8(m_b+m_c)}{f_{\chi_{b2}}f_{B_{c}}m_{B_{c}}^2
m_{\chi_{b2}}(4m_{\chi_{b2}}^2-\Delta)}e^{{m_{\chi_{b2}}^2}/{M^2}}
e^{{m_{B_{c}}^2}/{M'^2}}
\nonumber \\
&\times&\Bigg\{\int^{s_0}_{4m_b^2}ds
\int^{s_{0}^{'}}_{(m_b+m_c)^2}ds'
e^{\frac{-s}{M^2}}e^{\frac{-s'}{M'^2}}\rho_4(s,s',q^2)\Theta[L(s,s',q^2)]
+\widehat{{\cal B}}_{M^2}\widehat{{\cal
B}}_{M'^2}\Pi_4^{nonpert}(q^2)\Bigg\} \nonumber \\
&& {}
\end{eqnarray}
where $M^2$ and $M'^2$ are Borel mass parameters; and $s_0$ and
$s'_0$ are continuum thresholds in the initial and final channels.
Here, $\Theta$ is the step function and $L(s,s',q^2)$ is given by
\begin{equation}\label{L}
L(s,s',q^2)=s'y(1-x-y)-sxy+m_c^2(x+y-1)-m_b^2(x+y)+q^2x(1-x-y).
\end{equation}
The functions $\widehat{{\cal B}}_{M^2}\widehat{{\cal
B}}_{M'^2}\Pi_i^{nonpert}(q^2)$ are written as
\begin{equation}\label{exp}
\widehat{{\cal B}}_{M^2}\widehat{{\cal
B}}_{M'^2}\Pi_i^{nonpert}(q^2)=\langle\frac{\alpha_{s}GG}{\pi}\rangle\int^{1}_{0}dx
~~e^{\frac{{-m_{b}}^2(1+\frac{M^2x}{M'^2})+\frac{M^2x}{M'^2}(m_{c}^2-q^2x)}{M^2x(1+(-1+\frac{M^2}{M'^2})x)}}f_i(q^2),
\end{equation}
where $f_{i}(q^2)$ are very lengthy functions and
we do not present their explicit expressions  here.
\section{Numerical results}
To numerically analyze the sum rules obtained for the form
factors, we use the meson masses from PDG \cite{Olive}.  Considering the fact that the results of sum rules considerably depend on the quark masses, decay constants and gluon condensate, we use the values of 
these parameters from different sources. For the quark masses we take into account all the pole values and those obtained at $\overline{MS}$ scheme from the Refs. 
\cite{Abramowicz1,Abramowicz2,Narison,Chetyrkin1,Lee,Carrasco,Kiyo,Dehnadi,Ali,Chetyrkin2,Colangelo}. For $f_{B_{c}}$, we consider all the values predicted using different methods 
in Refs. \cite{fBc,fBcColquhoun,fBcSR,fBcpotmod}. In the case of $f_{\chi_{b2}}$, we use the  only value exists in the litreature, i.e.  $f_{\chi_{b2}}=(0.0122\pm0.0072)$
\cite{fKayb2}. For the gluon condensate, we also use its value from different sources \cite{Ioffe,Shifman1,Dominguez,Horsley,Chakraborty,Dominguez2,Geshkenbein,m021,m022} calculated via different approaches.

From the sum rules for the form factors it is also clear that they
contain extra four auxiliary parameters, namely the Borel parameters
$M^2$ and $M'^2$ as well as continuum thresholds $s_0$ and $s'_0$.
The general criteria is that the physical quantities like form
factors should be independent of these parameters. Therefore, we
need to determine their working regions such that the form factors
weakly depend on these parameters. To find the Borel windows, we
require that the higher states and continuum contributions are
sufficiently suppressed and the perturbative parts exceed the
nonperturbative contributions and the series of OPE converge. As a
result we get the windows: $ 14~ GeV^2 \leq M^2 \leq 20~ GeV^2 $
and $8~ GeV^2 \leq M'^2 \leq 12~ GeV^2 $. The continuum threshold
$s_{0}$ and $s'_0$ are not completely arbitrary but they are
related to the energy of the first excited states with the same
quantum numbers as the interpolating currents of the initial and
final channels. In this work the continuum thresholds are chosen
in the intervals $ 104~ GeV^2 \leq s_{0} \leq 108~ GeV^2$ and $
43~ GeV^2 \leq s'_{0} \leq 45~ GeV^2 $.

\begin{figure}[h]
\includegraphics[totalheight=7cm,width=8cm]{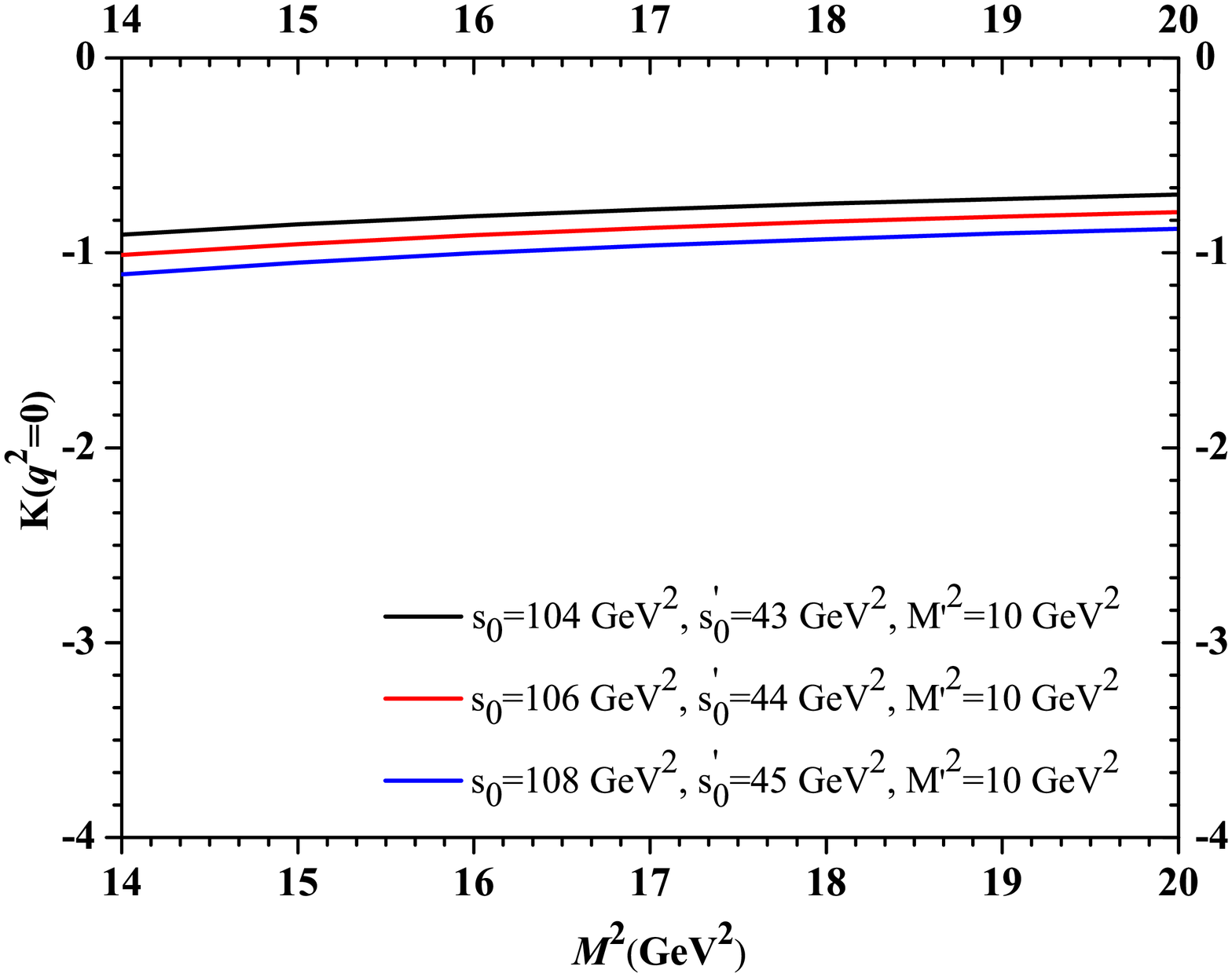}
\includegraphics[totalheight=7cm,width=8cm]{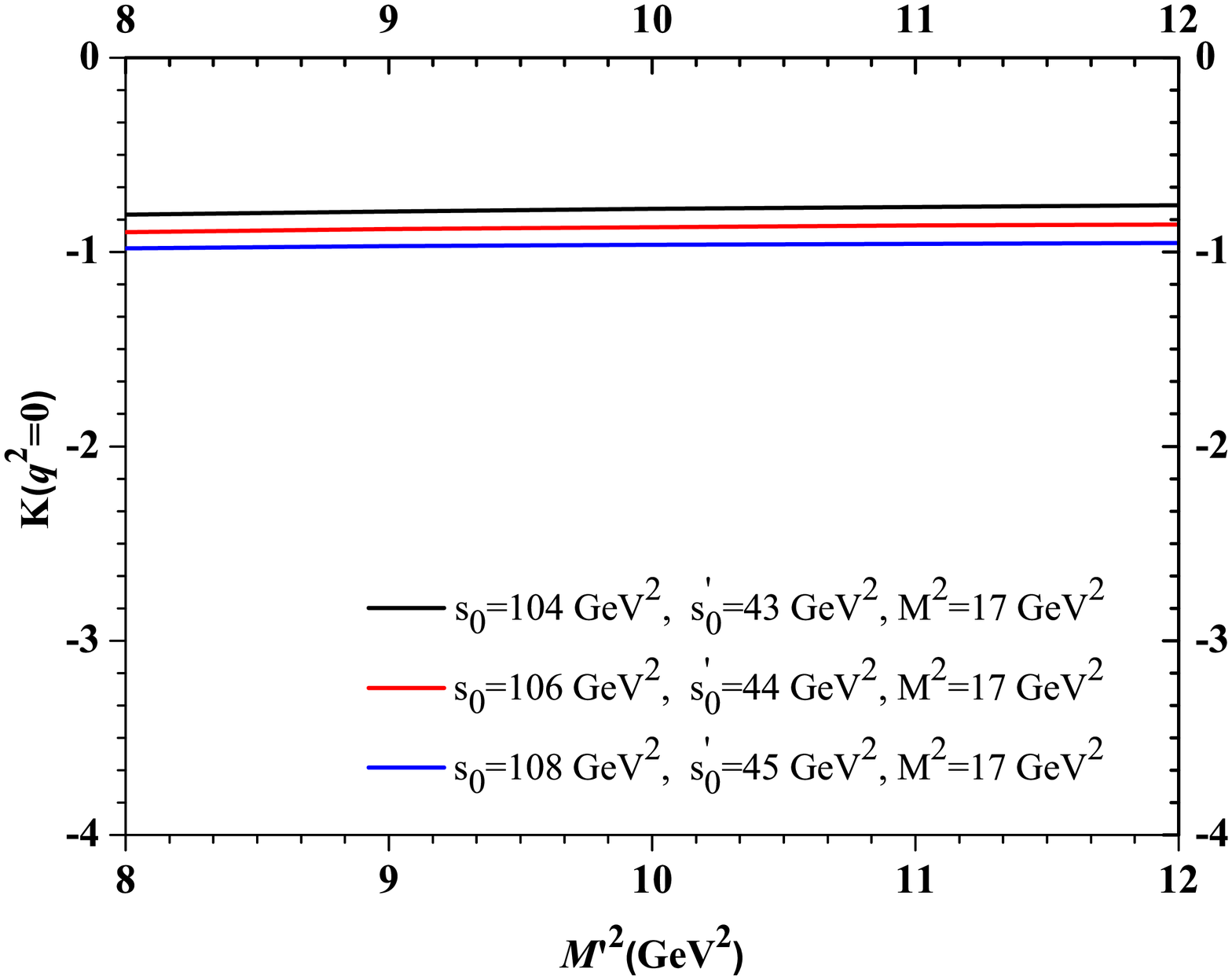}
\caption{Left: K($q^2=0$) as a function of the Borel mass
parameter $M^2$ at fixed values of $s_{0}$, $s_{0}^{\prime}$ and
$M'^2$. Right: K($q^2=0$) as a function of the Borel mass
parameter $M'^2$ at fixed values of $s_{0}$, $s_{0}^{\prime}$ and
$M^{2}$.} \label{KMsq}
\end{figure}

\begin{figure}[h]
\includegraphics[totalheight=7cm,width=8cm]{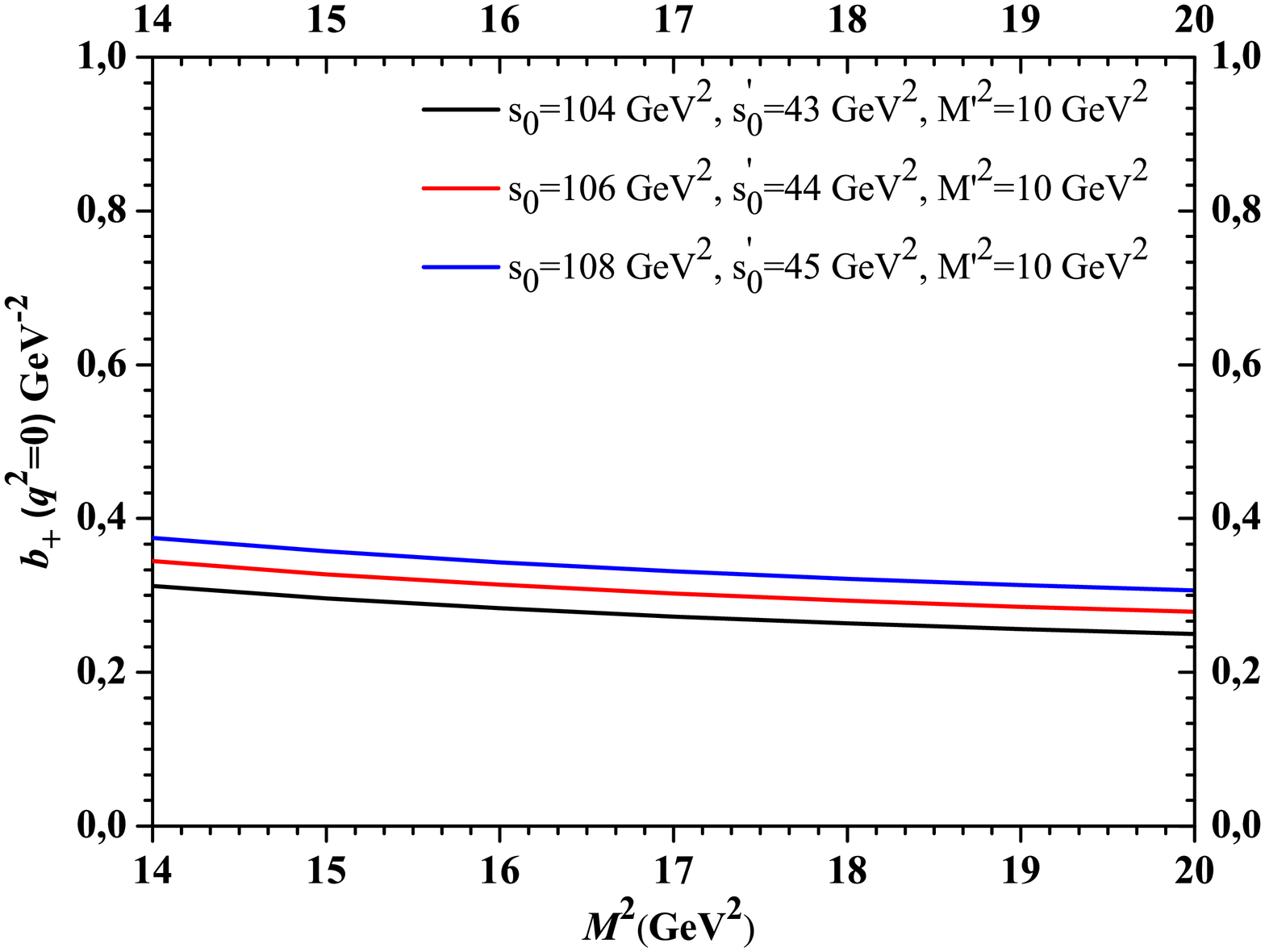}
\includegraphics[totalheight=7cm,width=8cm]{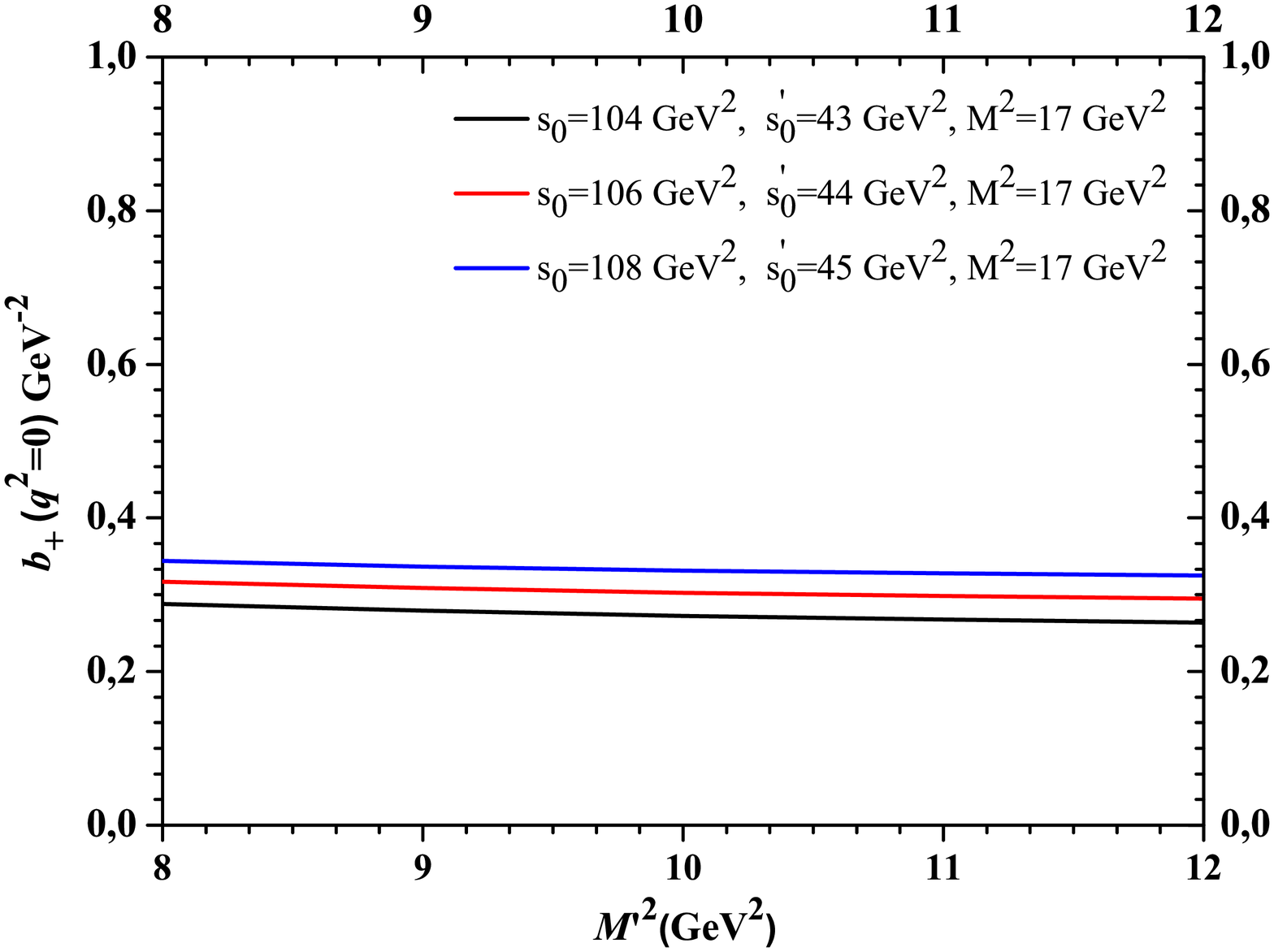}
\caption{Left: $b_{+}(q^2=0)$ as a function of the Borel mass
parameter $M^2$ at fixed values of $s_0$, $s_{0}^{\prime}$ and
$M'^2$. Right: $b_{+}(q^2=0)$ as a function of the Borel mass
parameter $M'^2$ at fixed values of $s_0$, $s_{0}^{\prime}$ and
$M^2$.} \label{bplusMsq}
\end{figure}

The dependence of form factors $K$ and $b_{+}$, as examples, on
Borel parameters $M^2$ and $M'^2$ at $q^2=0$ are plotted in
figures ~\ref{KMsq} and \ref{bplusMsq}.
From these figures we see that form factors show overall weak dependence on the  Borel mass parameters. The behaviors of the form
factors $K$, $b_{+}$, $b_{-}$ and $h$ in terms of $q^2$ are shown
in figures (3-6). In these figures, the red triangles show the QCD sum rule predictions, yellow-solid line denotes the prediction of fit function obtained using the central values of 
the input parameters and the green band shows the uncertainty due to errors of input parameters. Note that to obtain the central values, we consider the average values of input parameters discussed above, 
however, to calculate the uncertainties we consider all errors of these parameters from different sources previously quoted. As it is seen from these figures the sum rules
results are truncated at some points. Hence, to enlarge the region
to whole physical region we need to find some fit functions such
that their results coincide well with the QCD sum rules
predictions at reliable regions. For this reason we show the $q^2$
dependence of form factors including both the sum rules and fit
results in figures (3-6). Our numerical calculations reveal that
the following fit function well defines the form factors under
consideration:
\begin{equation}\label{fitfunc}
f(q^2)=f_0~
exp\Big[a\Big(\frac{q^2}{m_{\chi_{b2}}^2}\Big)+b\Big(\frac{q^2}{m_{\chi_{b2}}^2}\Big)^2
\Big],
\end{equation}
where the values of the parameters, $f_0$, $a$ and $b$ obtained
at $M^2=17~GeV^2$ and $M{'^2}=10~GeV^2$ for
$\chi_{b2}\rightarrow B_{c}\bar{l}{\nu}$ transition are presented
in table \ref{formfactors}.

Our final  purpose in this section is to obtain the decay width of
the $\chi_{b2}\rightarrow B_{c}\bar{l}{\nu}$ transition at all
lepton channels. The differential decay width for this transition
is obtained as
\begin{table}[h]
\renewcommand{\arraystretch}{1.5}
\addtolength{\arraycolsep}{3pt}
$$
\begin{array}{|c|c|c|c|}
\hline \hline
         &f_0 & a & b \\
\hline
  \mbox{$K (q^2)$}          &-0.871\pm0.279          &5.239\pm1.677    &-7.588\pm2.428 \\
  \hline
  \mbox{$b_{-} (q^2)$}      &-0.134\pm0.043~GeV^{-2}   &8.973\pm2.871    &56.462\pm18.068 \\
  \hline
  \mbox{$b_{+} (q^2)$}      &0.304\pm0.097~GeV^{-2}    &10.054\pm3.217  &53.922\pm17.255 \\
  \hline
  \mbox{$h (q^2)$}          &(-2.594\pm0.830)\times10^{-4}~GeV^{-2}   &5.224\pm1.672   &3.891\pm1.245 \\
                    \hline \hline
\end{array}
$$
\caption{Parameters appearing in the fit function of the form
factors.} \label{formfactors}
\renewcommand{\arraystretch}{1}
\addtolength{\arraycolsep}{-1.0pt}
\end{table}
\begin{eqnarray}\label{decaywidth}
\frac{d\Gamma}{dq^2}&=&\frac{G_{F}^2V_{cb}^2}{2^{10}3^2m_{\chi_{b2}}^7\pi^3q^6}(m_{l}^2-q^2)^2\Delta'^{3/2}\Bigg\{|b_{-}(q^2)|^2\Delta'm_{l}^2q^4
\nonumber\\
&+&|b_{+}(q^2)|^2\Delta'\Bigg[(m_{B_{c}}^2-m_{\chi_{b2}}^2)^2m_{l}^2+(m_{B_{c}}^2-m_{\chi_{b2}}^2)^2q^2-2(m_{B_{c}}^2+m_{\chi_{b2}}^2)q^4+q^6\Bigg]
\nonumber\\
&+&2Re[K(q^2)b^*_{+}(q^2)]\Delta'\Bigg[-q^4+m_{B_{c}}^2(m_{l}^2+q^2)-m_{\chi_{b2}}^2(m_{l}^2+q^2)\Bigg]
\nonumber\\
&-&2Re[b_{-}(q^2)b^*_{+}(q^2)]\Delta'm_{l}^2q^2(m^2_{B_{c}}-m^2_{\chi_{b2}})
\nonumber\\
&+&|K(q^2)|^2\Bigg[m_{B_{c}}^4(m_{l}^2+q^2)+m_{\chi_{b2}}^4(m_{l}^2+q^2)+q^4(m_{l}^2+q^2)-2m_{B_{c}}^2
\nonumber\\
&\times&(m_{\chi_{b2}}^2+q^2)(m_{l}^2+q^2)+m_{\chi_{b2}}^2q^2(m_{l}^2+5q^2)\Bigg]
\nonumber\\
&+&3|h(q^2)|^2\Delta'm_{\chi_{b2}}^2q^2(m_{l}^2+q^2)-2Re[K(q^2)b^*_{-}(q^2)]\Delta'm_{l}^2q^2\Bigg\}.
\end{eqnarray}
After performing integration over $q^2$ in Eq.(\ref{decaywidth})
in the interval $m_{l}^2\leq q^2 \leq(m_{\chi_{b2}}-m_{B_{c}})^2$,
we obtain the total decay widths as presented in
table~\ref{numdecaywidth} for different leptons.
\begin{table}[h]
\renewcommand{\arraystretch}{1.5}
\addtolength{\arraycolsep}{3pt}
$$
\begin{array}{|c|c|}
\hline \hline
  \mbox{}       &\Gamma(GeV)  \\
\hline
  \mbox{$\chi_{b2} \rightarrow B_{c}{\bar{e}}\nu_{e}$} &(1.054\pm0.506)\times 10^{-13} \\
  \hline
  \mbox{$\chi_{b2} \rightarrow B_{c}{\bar{\mu}}\nu_{\mu}$} &(1.041\pm0.500)\times 10^{-13}\\
  \hline
  \mbox{$\chi_{b2} \rightarrow B_{c}{\bar{\tau}}\nu_{\tau}$} &(2.398\pm1.175)\times 10^{-14}\\
\hline \hline
\end{array}
$$
\caption{Numerical results of decay widths at different lepton
channels.} \label{numdecaywidth}
\renewcommand{\arraystretch}{1}
\addtolength{\arraycolsep}{-1.0pt}
\end{table}
The errors belong to the uncertainties coming from the
determination of the working regions for auxiliary parameters as
well as those of the other input parameters. The orders of decay
rates at all lepton channels show that these transitions are
accessible at LHC in near future.

In summary, we have calculated the transition form factors for the
semileptonic $\chi_{b2}\rightarrow B_{c}\bar{l}{\nu}$ transition
using QCD sum rule technique. We took into account the two-gluon
condensate contributions as nonperturbative effects. We used these
form factors to estimate the order of decay widths at all lepton
channels. The order of decay width reveal that these transitions
can be seen at LHC in near future. Any comparison of the
experimental results with our predictions can provide us with
essential knowledge on the nature of the tensor $\chi_{b2}(1P)$
state.
\begin{figure}
\begin{center}
\includegraphics[totalheight=8cm,width=10cm]{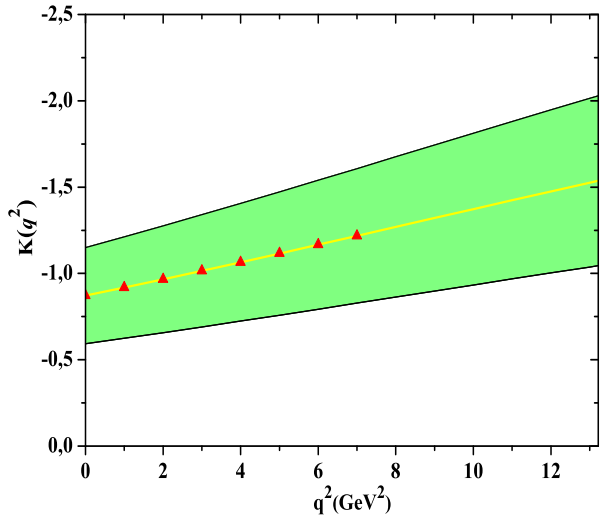}
\end{center}
\caption{ K($q^2$) as a function of $q^2$ at $s_{0}=106~GeV^2$,
$s_{0}^{\prime}=44~GeV^2$, $M^2=17~GeV^2$ and $M'^2=10~GeV^2$. The red triangles show the QCD sum rule predictions, yellow-solid line denotes the prediction of fit function obtained using the central values of 
the input parameters and the green band shows the uncertainty due to errors of input parameters. }
\label{Kq2}
\end{figure}
\begin{figure}
\begin{center}
\includegraphics[totalheight=8cm,width=10cm]{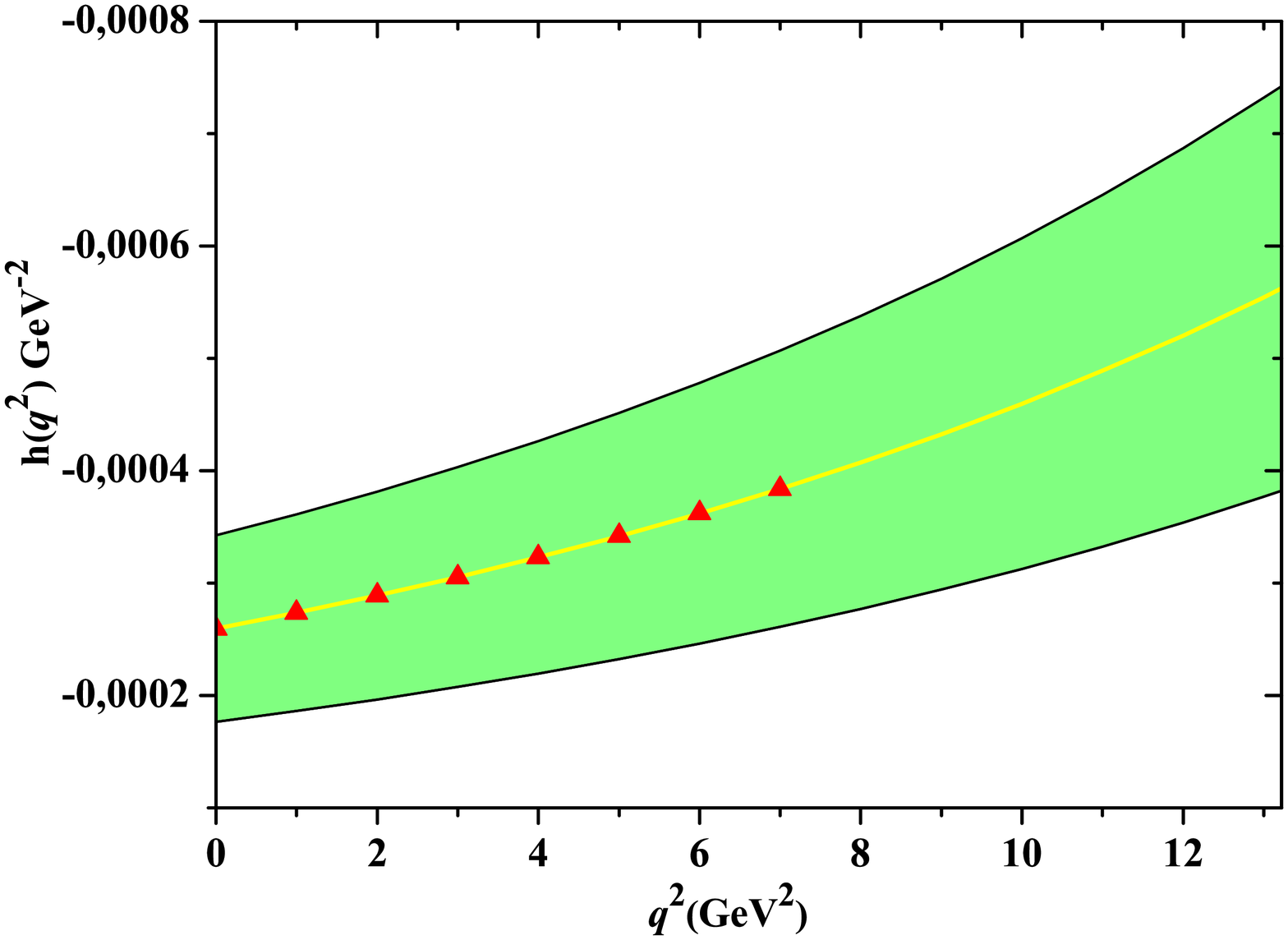}
\end{center}
\caption{The same as figure \ref{Kq2} but for  h($q^2$). }
\label{hq2}
\end{figure}
\begin{figure}
\begin{center}
\includegraphics[totalheight=8cm,width=10cm]{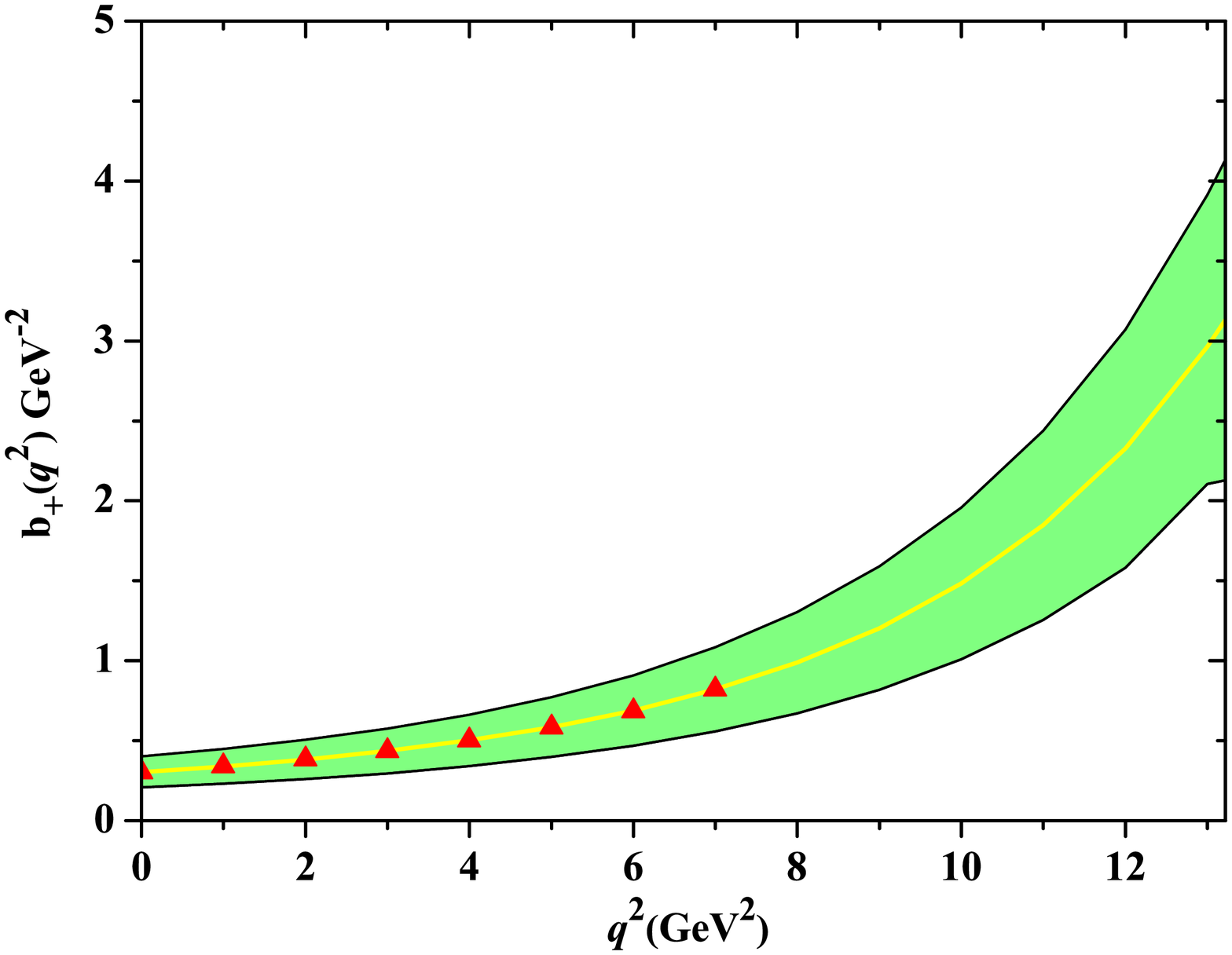}
\end{center}
\caption{The same as figure \ref{Kq2} but for $b_{+}(q^2)$. }
\label{bplusq2}
\end{figure}
\begin{figure}
\begin{center}
\includegraphics[totalheight=8cm,width=10cm]{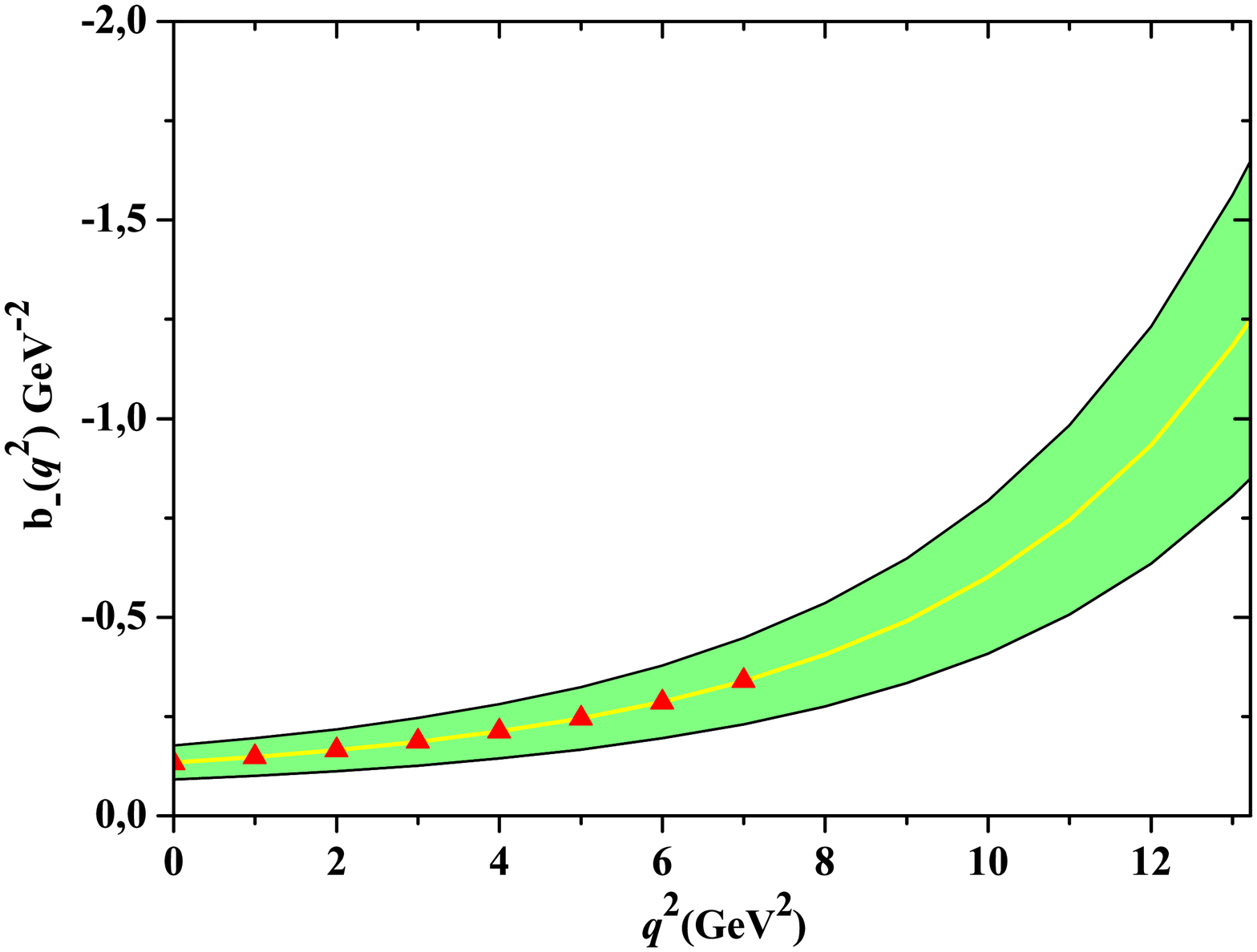}
\end{center}
\caption{The same as figure \ref{Kq2} but for $b_{-}(q^2)$. }
\label{bminusq2}
\end{figure}
%
%

\section{Conflict of Interests}

The
authors declare that there is no conflict of interest regarding
the publication of this paper.


\begin{thebibliography}{99}
%










\bibitem{Shifman1} M. A. Shifman, A. I. Vainshtein, V. I. Zakharov, ``QCD and resonance physics theoretical
foundations", Nucl. Phys. B \textbf{147}, 385 (1979).

\bibitem{Shifman2} M. A. Shifman, A. I. Vainshtein, V. I. Zakharov, ``QCD and resonance physics applications", Nucl. Phys. B \textbf{147}, 448 (1979).


\bibitem{Reinders85}L. J. Reinders, H. Rubinstein, S. Yazaki, ``Hadron properties from QCD sum rules", Phys. Rept. \textbf{127}, 1 (1985).
\bibitem{Olive} K. A. Olive et al. (Particle Data Group), ``Review of particle physics", Chin. Phys.
C \textbf{38}, 090001 (2014).




\bibitem{Abramowicz1} H. Abramowicz et al. [H1  and ZEUS Collaborations], ``Combination and QCD Analysis of Charm Production Cross Section Measurements in Deep-Inelastic ep Scattering at HERA'',
Eur.Phys.J. C \textbf{73}, 2311 (2013), arXiv:1211.1182 [hep-ex].


\bibitem{Abramowicz2} H. Abramowicz et al. [ZEUS Collaboration], ``Measurement of beauty and charm production in deep inelastic scattering at HERA and measurement of the beauty-quark mass'', 
JHEP \textbf{1409}, 127
(2014), arXiv:1405.6915 [hep-ex].

\bibitem{Narison} S. Narison, ``Summary on $m_{c,b}(m_{c,b})$ and precise
$f_{D(s),B(s)}$ from heavy-light QCD spectral sum rules",
Nucl. Phys. B - Proceedings Supplements \textbf{234}, 187 (2013),
16th Inter. Conf. in QCD, arXiv:1209.2925 [hep-ph].

\bibitem{Chetyrkin1} K. G. Chetyrkin et al., ``Charm and Bottom Quark Masses: an
Update", Phys. Rev. D \textbf{80}, 074010 (2009), arXiv:0907.2110
[hep-ph].

\bibitem{Lee} A. J. Lee  et al., ``Mass of the b-quark from lattice NRQCD
and lattice perturbation theory", Phys. Rev. D \textbf{87}, 074018
(2013), arXiv:1302.3739 [hep-lat].

\bibitem{Carrasco} N. Carrasco et al. [European Twisted Mass Collaboration], ``Up, down, strange and charm quark masses with $N_f = 2+1+1$ twisted mass lattice QCD'', Nucl.
Phys. B \textbf{887}, 19 (2014), arXiv:1403.4504 [hep-lat].

\bibitem{Kiyo} Y. Kiyo, G. Mishima and Y. Sumino, ``Determination of $m_c$ and $m_b$
from quarkonium 1S energy levels in perturbative QCD", Phys. Lett. B
\textbf{752}, 122 (2016), arXiv:1510.07072 [hep-ph].

\bibitem{Dehnadi} B. Dehnadi, A.H. Hoang and V. Mateu, ``Bottom and Charm Mass
Determinations with a Convergence Test",  JHEP \textbf{1508}, 155  (2015), arXiv:1504.07638 [hep-ph].

\bibitem{Ali} A. Ali, A.Y. Parkhomenko, A.V. Rusov,``Precise Calculation of the
Dilepton Invariant-Mass Spectrum and the Decay Rate in
$B^{\pm}\rightarrow \pi^{\pm} \mu^{+} \mu^{-}$ in the SM", Phys.
Rev. D \textbf{89}, 094021 (2014), arXiv:1312.2523 [hep-ph].

\bibitem{Chetyrkin2} K. G. Chetyrkin, M. Steinhauser, ``Short-Distance Mass of a Heavy
Quark at Order $\alpha^{3}_{s}$", Phys. Rev. Lett. \textbf{83}, 4001
(1999),

\bibitem{Colangelo} P. Colangelo,  A. Khodjamirian, ``QCD sum rules, a modern
perspective'',  At the Frontier of Particle Physics/Handbook of QCD,
edited by M. Shifman (World Scientific, Singapore), 3, 1495,
(2001), arXiv:hep-ph/0010175 .




\bibitem{fBc} M. J. Baker, J. Bordes, C. A. Dominguez, J. Penarrocha, K. Schilcher, ``B meson decay constants
$f_{B_c}$, $f_{B_s}$ and $f_B$ from QCD sum rules", JHEP
\textbf{1407}, 032 (2014) arXiv:1310.0941 [hep-ph].
%
\bibitem{fBcColquhoun} B. Colquhoun  et al., ``B-meson decay constants: a more complete picture from full lattice QCD", Phys. Rev. D \textbf{91}, 114509 (2015), arXiv:1503.05762 [hep-lat].
%
\bibitem{fBcSR} S. Narison, ``Improved $f_{D^*_{
(s)}}, f_{B^*_{(s)}}$ and $f_{B_{c}}$ from QCD Laplace sum rules",
Int. J. Mod. Phys. A \textbf{30}, 20, 1550116 (2015), arXiv:1404.6642
[hep-ph].
\bibitem{fBcpotmod} E. Bagan, H. G. Dosch, P. Gosdzinsky,
S. Narison and J. M. Richard, ``Hadrons with Charm and Beauty",
Z. Phys. C \textbf{64}, 57  (1994), arXiv:hep-ph/9403208.
%


\bibitem{fKayb2} T. M. Aliev, K. Azizi, M. Savci, ``Heavy $\chi_{Q_{2}}$ tensor mesons in QCD", Phys. Lett. B \textbf{690}, 164
(2010) arXiv:1002.2767 [hep-ph].





\bibitem{Ioffe} B. L. Ioffe, ``QCD at low energies", Prog. Part. Nucl. Phys. \textbf{56}, 232 (2006), arXiv:hep-ph/
0502148.
%
%
\bibitem{Dominguez} C. A. Dominguez, L. A. Hernandez and K. Schilcher, ``Determination of the gluon condensate from data in
the charm-quark region", JHEP \textbf{1507}, 110 (2015),
arXiv:1411.4500 [hep-ph].
%
\bibitem{Horsley} R. Horsley et al., ``Wilson loops to 20th order numerical stochastic
perturbation theory", Phys. Rev. D \textbf{86}, 054502 (2012),
arXiv:1205.1659 [hep-lat].
%
\bibitem{Chakraborty} B. Chakraborty et al., ``High-precision quark masses and QCD
coupling from $n_{f} = 4$ lattice QCD", Phys. Rev. D \textbf{91},
054508 (2015), arXiv:1408.4169 [hep-lat].
%
\bibitem{Dominguez2} C. A. Dominguez, L. A. Hernandez, K. Schilcher and H. Spiesberger,
``Chiral sum rules and vacuum condensates from tau-lepton decay
data", JHEP \textbf{1503}, 053 (2015), arXiv:1410.3779 [hep-ph].
%
\bibitem{Geshkenbein} B. V. Geshkenbein, ``Calculation of gluon and four-quark condensates
from the operator  expansion", Phys. Rev. D \textbf{70},
074027 (2004), arXiv:hep-ph/0309122.
%

\bibitem{m021} H. G. Dosch, M. Jamin, S. Narison, ``Baryon masses
and flavour symmetry breaking of chiral condensates", Phys. Lett.
B \textbf{220}, 251 (1989).


\bibitem{m022} V. M. Belyaev, B. L. Ioffe,
``Determination of baryon and baryonic masses from QCD sum sules.
Strange baryons", Sov. Phys. JETP \textbf{57}, 716 (1983).

\end{thebibliography}
\end{document}